# Testing Einstein's objection of 1927 to quantum mechanics opens the door to purely quantum communication

**Sergey A. Emelyanov**

*Ioffe Institute, 194021 St. Petersburg, Russia*

E-mail: sergey.emelyanov@mail.ioffe.ru

**Abstract**

Single electron may have wavefunction of a macroscopic lengthscale but cannot be detected in two places. At the Solvay Conference of 1927, Einstein argued that in a combination with Bohr's postulate about wavefunction as an exhaustive characteristics of electron, this fact implies nonlocality related to the collapse of a macroscopic wavefunction of single electron. This was his objection to Bohr's quantum mechanics (QM) in favor of de Broglie's quantum theory. We perform an experimental test in low-temperature state of matter where electrons have macroscopic orbit-like wavefunctions and surprisingly observe nonlocality of that kind. This nonlocality differs in principle from the well-known Einstein-Podolsky-Rosen (EPR) nonlocality because the former does not imply quantum entanglement and hence is beyond the no-communication theorem. Our observation is the first empirical justification of the choice in favor of QM in compare with de Broglie's theory. Moreover, now we can expand the current concept of reality and thus open the door to a realistic interpretation of quantum formalism. Ultimately, we thus come to a quantum model of Universe based on the standard QM.

# 1. Introduction

Although Einstein is one of the founding fathers of quantum mechanics (QM), this theory always seemed to him incomplete because it is does not explain physical world. Instead, it offers something like recipes to predict the outcomes of various experiments. Moreover, this theory implies an instantaneous process – the wavefunction collapse relevant even at a macroscopic lengthscale. Therefore, as such, this process is an existential threat to the Einstein's "kinematic" relativity.

At the Solvay Conference of 1927, as an argument against QM, Einstein argued that if, in accordance to QM, wavefunction is an exhaustive characteristics of electron, then there should be a nonlocality related to the collapse of a macroscopic wavefunction of single electron [1]. To clarify this nonlocality, Einstein proposed a *gedanken* experiment with single electron that has a macroscopic hemispherical wavefunction after the diffraction on a narrow slit. The point is according to QM such electron has no any preferable direction and may be detected at any (but only one) point of hemispherical screen. In this case, the wavefunction collapses immediately into this point to prevent a re-detection. Einstein regarded such collapse as a nonlocal process incompatible with his "kinematic" relativity and, in this context, the de Broglie's quantum theory seems preferable for him.

Historically, Bohr answered this objection when a one more objection related to nonlocality of QM was raised by Einstein, Podolsky and Rosen (EPR) in 1935. This new nonlocality was also related to the collapse of a macroscopic wavefunction but now it was the wavefunction of two (or more) entangled particles [2]. Actually, in both cases, Bohr's answers rest on the idea that quantum particle cannot be characterized by any method other than a direct measurement on it [3]. However, in violation of this rule, in the case of diffraction, we have only one point where the electron position may be regarded as measured (detection point) whereas, in the case of EPR, we determine the state of one particle through the measurements on another one. That is why Bohr believed that there is no nonlocality in both these cases and at that time his opinion was shared by most physicists.

Further progress in the field was related to the so-called Bell theorem that opens the door to the testing EPR nonlocality [4]. The best known are the experiments by Alain Aspect's group in the early 80s, which clearly show that EPR nonlocality does exists in accordance with QM [5]. This observation caused a wide public response far beyond the physical community. BBC radio station even released a series of programs where the leading physicists of that time shared their opinion about this observation because the public seemed that the Einstein's "kinematic" relativity appeared at stake though it is the basis of the current physical worldview [6]. However, a revolution in physics was avoided and the crucial point was that EPR nonlocality is related to quantum entanglement and, in this case, it can be proved mathematically that a nonlocal signaling is impossible even in principle [7]. Today this proof is called the no-communication theorem, according to which EPR nonlocality contradicts rather the spirit of "kinematic" relativity but does not lead to causality paradoxes [8].

Nevertheless, a few persons including physicist John Bell and philosopher Karl Popper supposed that EPR nonlocality is incompatible with "kinematic" relativity regardless of whether or not nonlocal signaling is possible [9-10]. They proposed to come back to the "dynamic" version of this theory by Hendrik Lorenz and Henri Poincare. The basic motive for such a radical proposal was that it is the only way that leaves us a chance to ever find a realistic version of quantum theory. Otherwise, we inevitably come to the so-called quantum dilemma that declares "either nonlocality or realism" and then, in the light of EPR nonlocality, we inevitably have to sacrifice quantum realism [11].

However, their proposal was rejected by physical community. First, most physicists believed that the problem of how to interpret relativity was already solved and cannot be the subject of a discussion at least because a lot of physical works are based precisely on the kinematic interpretation. Moreover, the kinematic version looks much more elegant than the dynamic one because the former does not require the so-called preferred reference frame always regarded as a vague medium (aether) permeating the entire Universe. Secondly, most physicists believed that the sacrificing quantum realism is not really a serious problem, at least because QM itself looks unrealistic even at the level of postulates.

As for realistic version of quantum theory, it is the so-called de Broglie–Bohm theory (dBB) which actually is a development by David Bohm of de Broglie's early ideas presented at the Solvay Conference [12-14]. One of the main differences between dBB theory and QM is that the former is completely deterministic and, in this regard, it is much closer to the classical physics where particles can always be characterized by spatial coordinates and therefore always have a definable trajectory. However, despite the fact that Bell himself and some other physicists associated quantum realism precisely with the dBB theory, Einstein nevertheless was rather skeptical concerning the determinism of this theory calling it literally "too cheap" [15].

Thus, the coexistence of two basic pillars of contemporary physics – QM and "kinematic" relativity – is based on the general belief that EPR nonlocality is the only possible type of quantum nonlocality. Perhaps that is the reason for a widespread misconception that nonlocality proposed by Einstein at the Solvay Conference of 1927 can always be reduced to EPR nonlocality (see, e.g., [16]). In this work we realize an experiment which, to our opinion, is truly of type proposed by Einstein because it is related to the collapse of a macroscopic wavefunction of single electron. We show that surprisingly such collapse does result in a deeper-than-EPR nonlocality incompatible with the kinematic relativity.

We perform our test in a modification of low-temperature state of matter known as the integer quantum Hall (IQH) state where electrons may have orbit-like wavefunctions of a macroscopic lengthscale. The test rests on the fact that the length of macroscopic electron jumps (if any) should be five orders longer than the electron's mean free path. Therefore, to demonstrate nonlocality, we need no tricky measurements with a very high temporal resolution. Instead, we only should demonstrate that a high enough number of photo-excited electrons appear in some well-defined regions that are at a macroscopic distance from laser spot, which by no means can be overcome by electrons through continuous spatial dynamics.

**Results**

**Concept.** We start with a comparison of Einstein's *gedanken* experiment with our experiment in IQH system. The former is shown schematically in Fig. 1 (left panel), where, for simplicity, we assume that the electron wavefunction after diffraction is one-dimensional. Here *B* is the point, in which electron is detected on the screen and *A* is a distant region where the wavefunction is nonzero just before the collapse. If wavefunction is an exhaustive characteristics of electron, then electron may be regarded as being in the region *A* just before the collapse providing thus a deeper-than-EPR nonlocality at the collapse (this fact was noted also in [17]).

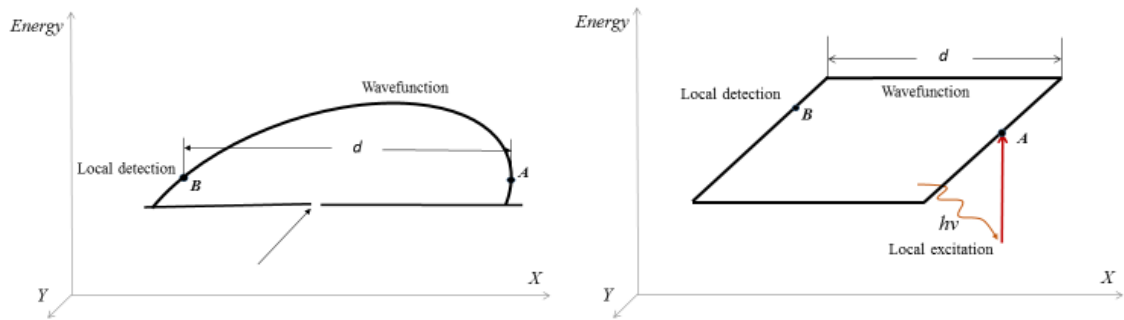

Fig.1. Thought experiments with nonlocality related to the collapse of a macroscopic wavefunction of single electron.
Left panel – Einstein's thought experiment with the electron diffraction on a narrow slit with a subsequent detection of this electron at the point *B* on a hemispherical screen.
Right panel – thought experiment with a strictly local photo-excitation of electron at the point *A* in a square macro-orbit of IQH system with a subsequent detection of this electron at the point *B* through the scattering on immobile point-like charged centers.

In the case of IQH system, electron also has a macroscopic wavefunction which, however, is stationary and belongs to the so-called extended (or current-carrying) state. Actually it is a quasi-one-dimensional macro-orbit extended along the perimeter of a macroscopic (in two dimensions) IQH system [18-21]. Today the existence of such extended states is out of doubts because they are responsible for the

IQH effect itself, the discovery of which by Klaus von Klitzing was awarded the Nobel Prize in 1985 [22]. Thus, if, for definiteness, we take a square IQH system, then each macro-orbit should also be a square. One of such orbits is shown in Fig. 1 (right panel). The similarity of both situations in Fig. 1 becomes more clear if we take into account that macro-orbits are typically immersed in the system of immobile point-like charged scatterers. In this case, the electron may be detected (or scattered) in any (but only one) point of the orbit precisely like after the diffraction, electron may be detected (or scattered) in any (but only one) point of hemispherical screen. According to QM, to prevent a re-detection, in both cases macroscopic wavefunction collapses instantaneously into the point of detection.

As noted, in Einstein's *gedanken* experiment, electron is regarded to be in the region *A* just before the collapse because here the wavefunction, as an exhaustive characteristics of electron, is nonzero. But in the case of IQH system, we are even capable of proving experimentally that electron was really in the region *A* before the collapse and was really in the point *B* a time *t* after the collapse, where *t* is the electron lifetime in the orbit. To this aim, electron should be excited into the orbit due to a strictly local interaction in the region *A*, for example, with a photon. In this case, to achieve nonlocality, the lifetime *t* should be less than $d/c$ where $d$ is the system length, $c$ is the speed of light. Taking into account that in a real IQH systems, the scattering time may be of the order of 3ps whereas *d* may be about 1cm, we obtain that *t* may be one order shorter than $d/c$.

In the case of square macro-orbit and homogeneous distribution of point-like charged scatterers, the photo-excitation of electron close to the right side of the orbit, with a probability of one quarter will result in the detection of the electron close to the left side. This means with the probability of one quarter the electron overcomes the distance of no less than *d* without having visited any intermediate points. In fact, this is a discrete spatial dynamics of electron, which definitely is beyond the kinematic relativity. Moreover, the probability of such macro-jump can easily be increased, for example, through the increasing of local density of scatterers near the left side of the orbit.

In the above reasoning, we use the characteristic features of QM, which follow from the fact that the wavefunction is an exhaustive characteristics of electron. One of these features is that electron cannot occupy a part of quantum orbit but only the whole orbit even if the orbit is macroscopic. That is why QM has no such concept as the rate of filling (or release) of quantum orbit. These processes are instantaneous. The other feature (correlated with the first one) is that the electron has no any definable spatial position in the orbit without measurement and behaves as if it occupies all possible spatial positions simultaneously. As a result, although electron appears in the orbit due to a local excitation in the region *A*, we nevertheless suppose that it occupies instantly the whole orbit and may be detected equiprobably in any point of the orbit.

Of course, the above features make QM non-realistic at least in terms of the relativistic insight of reality limited by the four-dimensional spacetime. As a matter of fact, that is the main reason why Einstein believed that QM is incomplete and why he proposed a single-electron nonlocality as an objection to QM and was absolutely sure (like most physicists) in the ultimate success of his kinematic relativity when clashing with QM. Nowadays, however, this success does not look so evident at least because QM is generally recognized as the most successful theory in the history of science.

**Material.** One of the problems of our experiment is that all macro-orbits are actually edging states at least in ordinary IQH system [23]. Therefore, only a small fraction of electrons may potentially be involved. To solve this problem, consider first an IQH systems in more details. As a rule, the starting material for such system is a two-dimensional semiconductor structure where the electron motion in one direction (*Z*) is quantized because the width of conducting layer in this direction (a few tens of nm) is less than the electrons' mean free path. The energy spectrum in this direction is thus a series of quantum levels.

Next, there should also be a strong perpendicular magnetic field *B* so that the electron motion becomes quantized in all directions and each quantum level gives rise to a series of sublevels called Landau levels. Here most electrons are in the flat cyclotron orbits which typically are microscopic. Finally, in any real IQH system, there is a band bending that gives rise to an electric field perpendicular to the system edges. This field occurs only close to the edges within a narrow strip of the order of cyclotron radius. As a result, a crossed electric and magnetic field emerges there and, as it was first shown by Bertrand Halperin, this field may lift Landau degeneracy and give rise to the extended states capable of providing macroscopic currents [24]. That is why, despite of a small absolute number of extended states, these states are detectable in magneto-transport measurements.

However, as it was shown in [25], a number of macro-orbits in the system interior may be reached if IQH system has an asymmetry of confining potential due to asymmetric ionization of atoms near the interfaces. This asymmetry results in a transverse electric field that may be as high as of the order of $10^5$V/cm [26]. It is the so-called "built-in" field that usually manifests itself in spin-related effects such as the Rashba effect [27-28]. However, if external magnetic field has both quantizing and in-plane components, then all electrons appear in a crossed electric and magnetic field and macro-orbits may occur not only near the edges but also in the system interior [29-30].

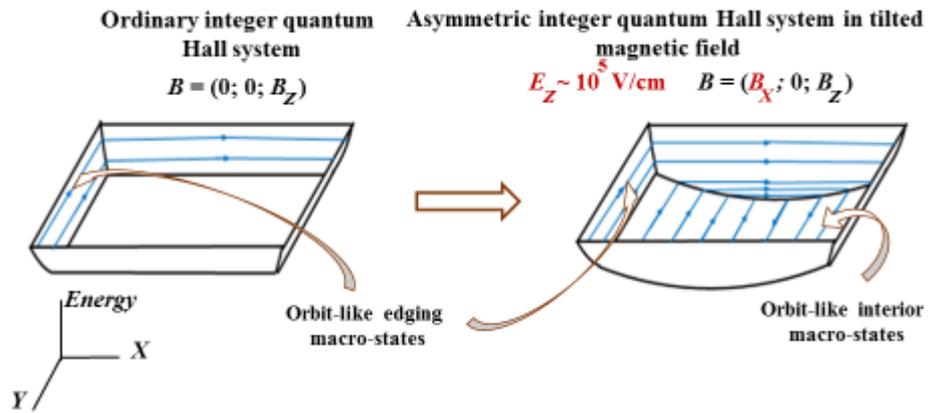

Fig. 2. Energy diagram of Landau level. Left panel – ordinary IQH system. Right panel – asymmetric IQH system in tilted quantizing magnetic field.

Energy diagram of a Landau level in ordinary IQH system is shown in Fig. 2 (left panel). The same diagram for asymmetric IQH system in tilted quantizing magnetic field is shown in Fig.2 (right panel). It is seen that in the former case, Landau level resembles rather an "energy saucer" with a flat bottom and extremely narrow walls that contain macro-orbits. But in the latter case the bottom of "energy saucer" is concave in the $X$ direction. As a result, macro-orbits cross the whole system along the $Y$ axis being closed through edging states parallel to the $X$ axis. Thus, each Landau level consists of a number of macro-orbits and bearing in mind the spatial distribution of such orbits, our test looks like this.

**Experiment.** We take a single quantum well structure where a few Landau levels are always below the Fermi level and hence filled with electrons even at liquid helium temperatures. In this case, the electron transitions occur between the highest filled level (N – 1) and the lowest vacant level (N). As a rule, excitation is provided by terahertz laser radiation.

Typical IQH system is 21mm in length ($X$-axis) and 17mm in width ($Y$-axis). It is covered with a non-transparent mask that has six identical windows symmetrical with respect to the system center (Fig. 3). Each window is 4mm in length and 5mm in width. Taking into account the spatial distribution of macro-orbits, the windows can be divided into two families, each of which is crossed by the same macro-orbits. The first family is from the windows №1 to the windows №4, and the second family is from the windows No. 5 to the window No. 6. Then, if single-electron nonlocality does exist, we should observe the following non-trivial effect. When electrons are excited through any windows from No. 1 to No. 4, they should be detected not only in the illuminated region, but in *all* regions of this family and moreover the number of electrons in these regions should be almost the same *regardless* of which window was really open.

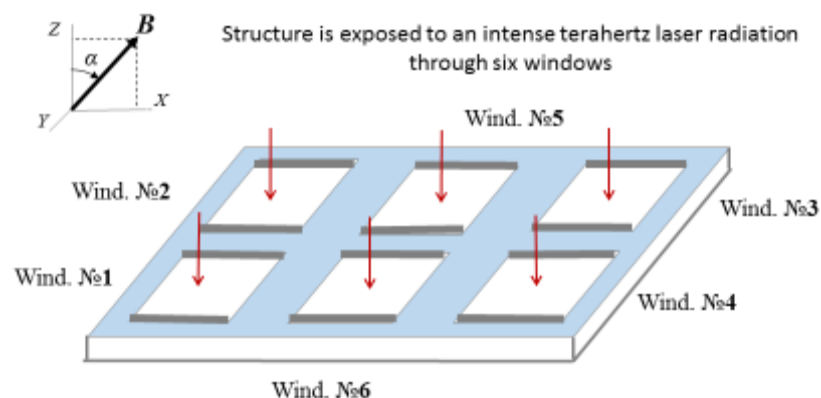

Fig. 3. Experimental scheme. The structure is covered with a non-transparent mask with six identical windows through which terahertz laser radiation enters the structure. Each region is equipped with a pair of ohmic contacts to measure local current along the *Y*-axis. External magnetic field (*B*) may be tilted in the *XZ* plane by an arbitrary angle.

Similarly, if we illuminate either the window No. 5 or the window No. 6, then almost the same number of photo-electrons should appear in both regions regardless of which window is really open. But any manipulations with the windows of one family should not affect the regions of another family. Finally, if there is no single-electron nonlocality at all, then there should be no photo-electrons in any regions in the dark because the electrons' mean free path is as short as about 0.1μm.

To avoid any ambiguity, we perform our test in two regimes that differ from each other only in the direction of magnetic field switched *in situ*. In the first regime (regime I), magnetic field is parallel to the structure so that there is no Landau quantization. In the second regime (regime II), quantizing magnetic field is slightly tilted and macro-orbits should exist. In each regime, windows are consecutively closed and we measure the relative number of excited electrons in each working region. The results are presented as a bar chart where the height of the column is proportional to the number of photo-electrons (Fig. 4). Yellow color means that this window was open during the measurement, blue color – it was closed. Left column – regime I, right column – regime II.

It is seen that the system behavior is quite trivial in the regime I. Indeed, photo-electrons emerge only if the window is open and the number of electrons is almost the same in any illuminated region. This means all working regions behave independently of each other. By contrast, in the regime II, our non-trivial scenario is fully confirmed. Photo-electrons emerge not only in the illuminated regions but also in the regions in the dark. Moreover, if some regions belong to the same family, then the number of electrons in each of them is almost the same regardless of which of them is really exposed to the laser radiation. Finally, as expected, if regions are from different families, then the illumination of one of them do not lead to any electrons in another region.

**Discussion**

Thus, our test clearly indicates that we are truly facing macroscopic electron jumps that may be regarded as a discrete spatial dynamics (DPD) of electrons at a macroscopic distance. Such dynamics clearly implies a nonlocal signaling and therefore by no means is compatible with the kinematic relativity. However, although this observation is consistent with QM, it nevertheless seems incredible.

**Quantum view of the Universe.** The reason is the same as in the days of Bell and Popper: kinematic relativity is the basis of our current view of physical world and moreover it is the basis of a great number of works especially in the field of cosmology. However, it may be that our observations are capable of providing the basis for a deeper quantum view of physical world. The point is the DPD nonlocality brings us back to the concept of absolute simultaneity. But now this is not a classical concept based on the meaningless idea of an infinite speed. Instead, now we have a quantum concept of absolute simultaneity based on the well-tested DPD nonlocality compatible with the dynamic relativity.

**Closed windows: No. 6**

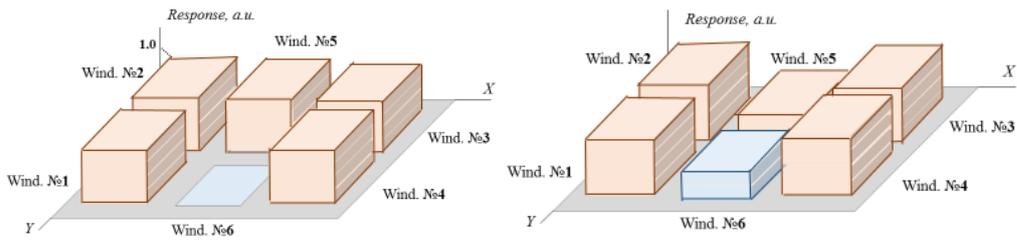

**Closed windows: No. 6, 5**

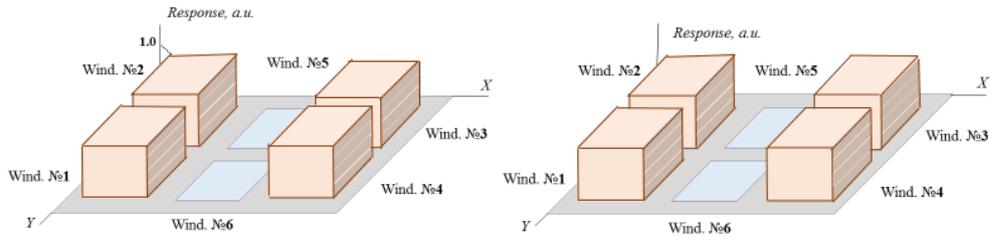

**Closed windows: No. 6, 5, 4**

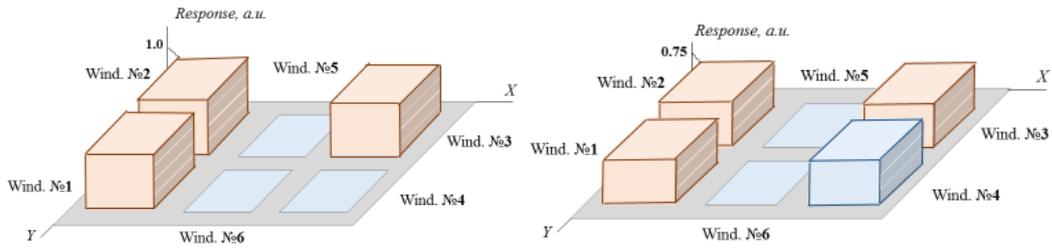

**Closed windows: No. 6, 5, 4, 3**

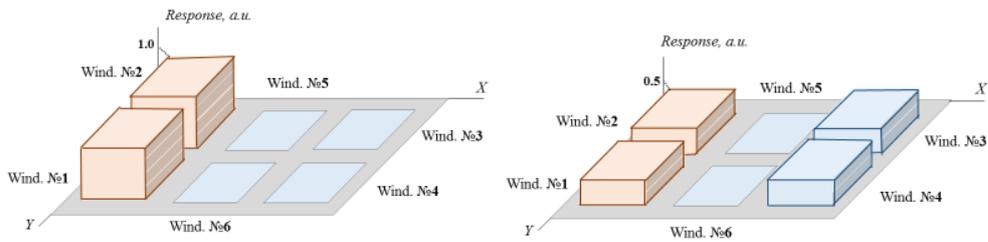

**Closed windows: No. 6, 5, 4, 3, 2**

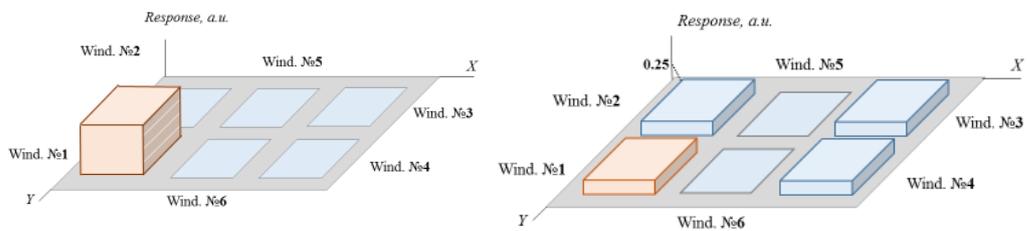

Fig. 4. Bar charts of normalized local photo-responses under different configurations of local photo-excitation. External magnetic field is 4.8T. Left column – regime I ($\alpha = 90°$); right column – regime II ($\alpha = 8°$).

Having in mind the quantum absolute simultaneity, we can now take a fresh look at the quantum model of undivided Universe proposed by David Bohm and Basil Hiley in the early 90s [31]. In fact, the main problem of this model is that it is trying to reconcile EPR nonlocality with the relativistic concept of spacetime though actually it is hardly possible, at least within a realistic approach. As a result, this model remains almost unclaimed by physicists. However, if EPR nonlocality is replaced by DPD nonlocality, then surprisingly we can easily replace kinematic relativity by the dynamic one because the problem of preferred reference frame now can be solved without the vague notion "aether". Indeed, through the quantum concept of absolute simultaneity, we can easily introduce both true time and true length and then, for any object regarded separately from undivided Universe, the rest of such Universe will naturally be the preferred reference frame with true time and true length.

**Quantum view of reality.** However, at first glance, our approach leads to a contradiction related to the fact that our quantum model of Universe rests not on the dBB theory but on QM known to be unrealistic. However, the point is the replacement of kinematic relativity by the dynamic one surprisingly opens the door to a realistic insight of the quantum formalism itself. Indeed, according to this formalism, a particle may be in a superposition of pure states and only the so-called measurement forces the particle to "choose" one of the pure states. That is why quantum formalism appeals to a multidimensional Hilbert space, i.e. to the space of quantum states, rather than to the four-dimensional spacetime as the framework of relativistic reality [32]. Einstein himself emphasized this problem at the Solvay Conference of 1927 where the return of QM to the framework of "real" space he regarded as the main task of this theory and at that time most participants shared his opinion. Today, however, it becomes more and more clear that this task is actually unsolvable because the appeal to Hilbert space lies in the heart of QM.

But now one can propose an alternative way of how to solve the problem of quantum realism. The point is insofar as we come back to the dynamic version of relativity where time is not closely linked with spatial coordinates, we can expand the concept of reality beyond the frames dictated by relativity. We can imagine that reality consists of two parts: observable three-dimensional reality and unobservable reality definable in multidimensional Hilbert space.

Then, if particle is in a superposed state, it nevertheless remains to be an element of reality. What we call measurement appears thus to be a physical process when particle becomes an integral part of macroscopic body (apparatus) and is forced to "choose" one of its eigenstates in order to transit from unobservable reality to the observable one. Which of quantum states may be "chosen" is determined by the macroscopic body. It that case, the presence of an observer is clearly optional and we no more need to attract the mystical influence of the consciousness of living beings on the "choice" of quantum particle. Actually, this fact was stressed long time ago by Landau and Lifshitz in their famous course of quantum mechanics where they wrote literally the following: *"… we emphasize that, in speaking of 'performing a measurement', we refer to the interaction of an electron with a classical 'apparatus', which in no way presupposes the presence of an external observer"* [33].

Thus, in the light of a deeper-that-EPR nonlocality, there is a chance to make QM completely realistic like the other fundamental physical theories. It would be precisely what Einstein required from QM throughout all his life but nowadays his requirement is regrettably the reason for reproaches of unhealthy conservatism or even stubbornness [34]. However, Einstein was not the only person concerned about the lack of realism in QM. His concerns were shared, for example, by John Bell, whose sympathies for the dBB theory are related primarily to a realistic character of this theory [9]. Finally, the deep experiences and doubts about the lack of a worldview in QM were expressed by Richard Feynman and moreover in a rather poetic manner [35]. But now perhaps a light has dawned at the end of the tunnel in this matter.

**Conclusion**

We have realized an experiment of such type as was proposed by Einstein as an objection to Bohr's QM at the Solvay Conference of 1927. There Einstein argued that QM predicts nonlocality related to the collapse of a macroscopic wavefunction of single electron and therefore is incompatible with his kinematic version of relativity. To test this point, we have used a modification of so-called quantum Hall state of matter where electrons possess macroscopic orbit-like wavefunction. As a result, we have truly observed nonlocality of that type. The nonlocality is unrelated to quantum entanglement and actually is a discrete spatial dynamics of individual electrons. This observation makes it possible to empirically justify the choice in favor of QM in comparison with de Broglie's quantum theory as well as to expand the concept of reality in such a way as to interpret realistically quantum formalism and thereby to develop a quantum concept of Universe on the basis of the Bohm-Hiley concept of undivided Universe.

**Method**

The main problem of our method is how to provide a counter of photo-electrons in the working regions. Since we are dealing with macro-orbits known to be responsible for macroscopic currents, one would expect that the local scattering of electron in the orbit results in a local current in the $Y$ direction precisely where the scattering occurred. Thus, the photo-induced local currents may be a counter of photo-electrons in the working regions. Phenomenologically, the reason for such currents under the excitation by terahertz laser radiation is an inner system asymmetry characterized by the following vector product: $B_x \times E_{built-in}$. Actually, this is the so-called photo-voltaic effect [36]. But, as a rule, the effect is observed in a system of Bloch electrons which are not spatially separated and therefore the effect should not depend on spatial coordinates. In our case, however, we are dealing with the system of spatially separated electrons and therefore local currents may be a function of spatial coordinates.

To measure local photo-induced currents along the $Y$ axis, each working region is equipped with a pair of ohmic contacts (0.5mm wide) parallel to the $X$ axis. Each contact lies not less than 1.5 mm away from the edges. The distance between the adjacent regions is no less than 3mm. The structures are grown by the method of molecular beam epitaxy (MBE). They are semi-metallic single quantum wells of type GaSb-InAs-GaSb with the well width of 15 nm. In structures of that type, the conduction band of InAs overlaps the valence band of GaSb so that the electron density is as high as of the order of $2 \times 10^{12}$ cm$^{-2}$ even at low working temperature (about 1.9K). However, since the overlapping may result in the so-called hybridization of electrons, the conducting layer of InAs is sandwiched between two very thin layers of AlSb (3nm each). To provide built-in field, the conducting layer is removed from the surface by 20 nm whereas the so-called surface potential is known to penetrate into the structure to a depth of about 100 nm [37].

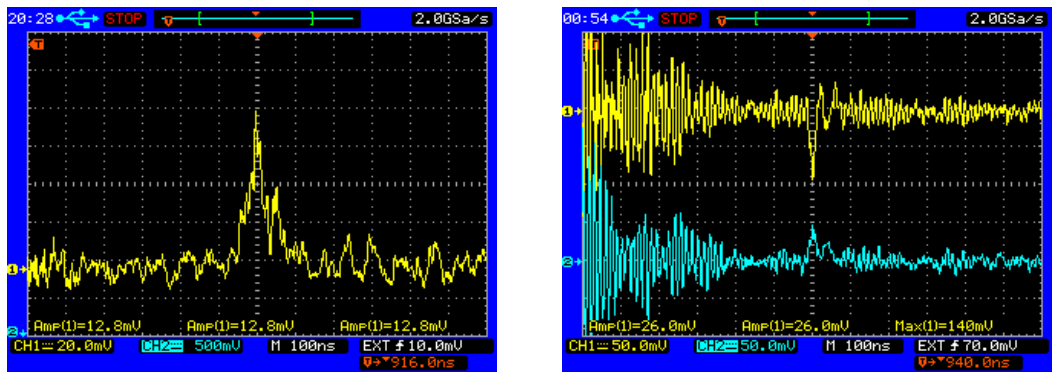

Fig. 5. Left panel – Typical track of terahertz laser pulse. Timescale is 100ns/div. Right panel – typical track under a synchronous detection of photo-responses in both the illuminated region No. 1 (upper track) and the region No. 3 in the dark (lower track), α = − 8˚, B = 4.8T. Timescale is 100ns/div.

As a source of radiation, we use pulsed terahertz ammonia laser optically pumped by $CO_2$ laser. The wavelength of terahertz radiation is 90.6μm (photon energy 13.7meV). Since photovoltaic effect is, as a rule, relatively small, we use the method of intense terahertz laser spectroscopy [38]. In our case, the intensity of radiation is about 200W/cm$^2$. To avoid a remarkable heating by this radiation, we use short laser pulses of about 40ns. Typical track of the laser pulses is shown in Fig. 5 (left panel).

An advantage of our time-resolved measurements is that we always can be sure that photo-excited electrons far from laser spot (if any) cannot be related to a slow diffusion from the spot as then they should distort the kinetics of photo-response. Fig. 5 (right panel) shows typical tracks of local currents in two regions belonged to the same family, one of which is illuminated (yellow track) whereas another one is in the dark (blue track). It is clearly seen that the kinetics of both signals is not distorted and there is no a delay between them despite the fact that the region in the dark is removed from the laser spot by a macroscopic distance.

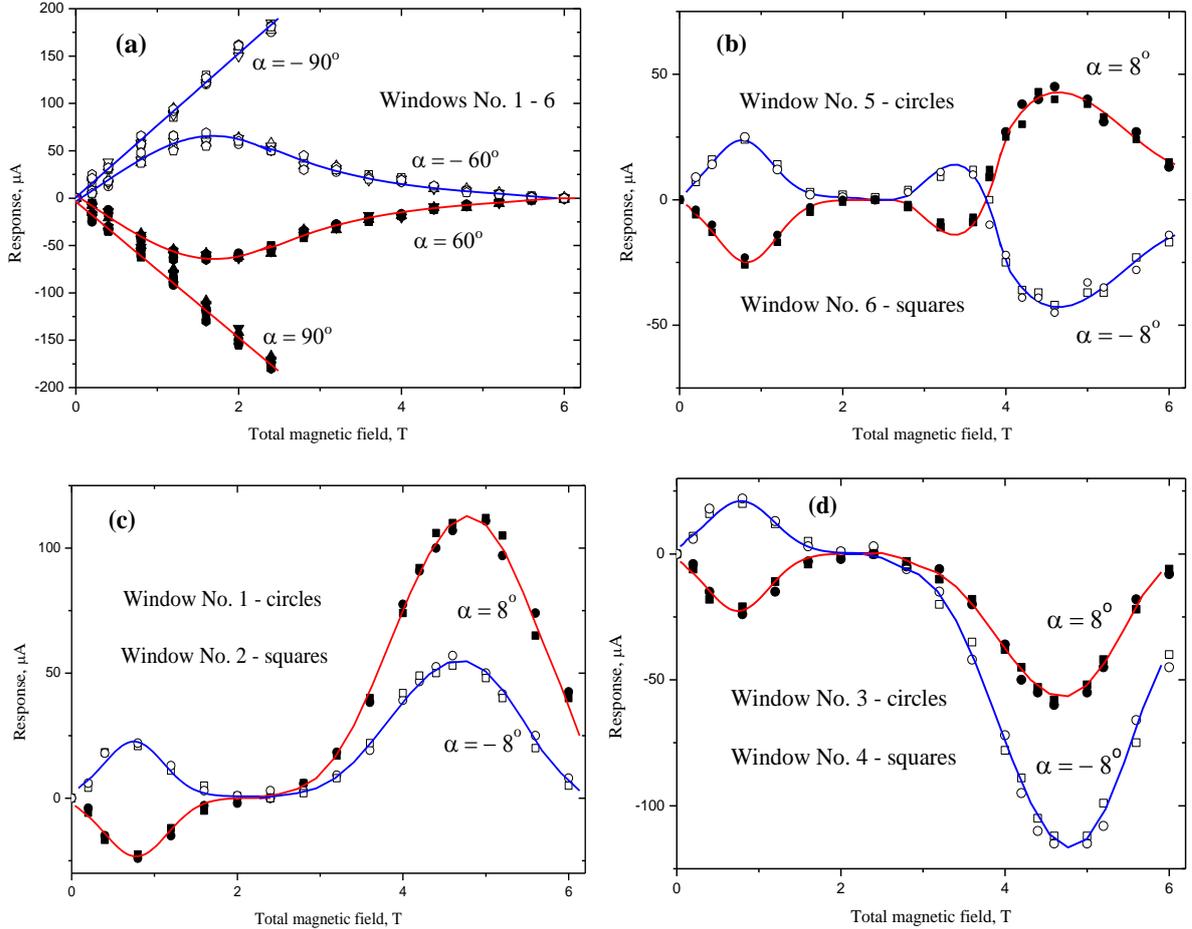

Fig. 6. (a) – the dependence of six local photocurrents on the total magnetic field under the full illumination in the regime I Open symbols – negative in-plane magnetic field (blue lines); solid symbols – positive in-plane magnetic field (red lines).
(b, c, d) – the dependence of local photocurrents under the full illumination in the regime II: (b) – regions No.5 and No. 6, (c) – regions No. 1 and No. 2, (d) – regions No. 3 and No. 4. Solid lines are a guide for the eyes.

To check the applicability of our counter, we measure local currents in all working regions in the regime of Bloch electrons as well as in the regime of macro-orbits under the full illumination. In the former case, we measure local currents when magnetic field is parallel to the well plane and there is no Landau quantization. The results for two opposite directions of $B_x$ ($\alpha = 90°$ and $\alpha = -90°$) are shown in Fig. 6a. To compare, we also show the results at $\alpha = 60°$ and $\alpha = -60°$, i.e. when quantizing component is nonzero. It is seen that here the behavior of local responses is quite trivial. They are the same in all regions and when $\alpha = \pm 90°$ they are almost proportional to the magnetic field. However, when $\alpha = \pm 60°$, local currents disappear at high magnetic field because of the Landau quantization.

However, the system behavior differs in principle in the regime of macro-orbits. Here the same currents in working regions are observed only far from the cyclotron resonance ($B < 3$T). However, close to resonance, local currents are the same only if their regions have the same $X$ coordinates. We present the results separately for three pairs of working regions with the same $X$ coordinates (Fig. 6b, c, d). It is seen that local currents differ from each other not only in their absolute value but also in their sign. Thus, as expected, the sensitivity of our counter is a strong function of the $X$ coordinate. That is why, to compare local currents, each of them is always normalized on the current observed in this region under the full illumination.

Finally, it should be noted that the system of spatially separated macro-orbits is actually a highly-correlated quantum macro-system. As a result, the basic symmetry relations meet only for the system as a whole, but may not meet for local regions. This fact is clearly seen from Fig. 6b, c, d where changing the sign of magnetic field does not lead to the changing the sign of local currents if the regions is asymmetric with respect to the system center in the $X$ direction. This means the behavior of such regions cannot be analyzed separately from the whole system.

**Finding**

This research received no external funding.

**Competing interests**

**Acknowledgements**

The author is grateful to Prof. Sergey Ivanov (Ioffe Institute) for the MBE samples as well as to Prof. Raymond Chiao (UC at Merced) for useful comments on the experiment.

**References**

1. Bacciagaluppi G., Valentini A., *Quantum Mechanics at the Crossroads. Reconsidering the 1927 Solvay Conference*, Cambridge University Press, Cambridge (2009).
2. Einstein A., Podolsky B., Rosen N., *Can quantum-mechanical description of physical reality be considered complete?* Phys. Rev. **47** (1935) 777-780.
3. Bohr N., *Can quantum-mechanical description of physical reality be considered complete?* Phys. Rev. **48** (1935) 696-702.
4. Bell J. S., *On the Einstein-Podolsky-Rosen paradox*, Physics, **1** (1964) 195-200.
5. Aspect A., Grangier P., Roger G. *Experimental tests of realistic local theories via Bell's theorem*. Phys. Rev. Lett., **47** (1981) 460-463.
6. Davies P. C. W. and Brown J. R. (eds.), *The ghost in the atom* Cambridge University Press (1986).
7. Eberhard P.H., Bell's theorem and the different concepts of locality. *Il Nuovo Cimento* 46B, 392–419 (1978).
8. Ghirardi G. C., Grassi R., Rimini R., Weber A., *Experiments of the EPR type involving CP-violation do not allow faster-than-light communication between distant observers* Europhysics Letters **6** (1988) 95-100.
9. Bell J. S., *Speakable and Unspeakable in Quantum Mechanics*, Cambridge University Press (1987).
10. Popper K., *A critical note on the greatest days of quantum theory*, Found. Phys. **12** (1982) 971-976.
11. Whitaker M. A. *Einstein, Bohr and the quantum dilemma*, Cambridge University Press (2006).
12. Bohm D. J., *A suggested interpretation of the quantum theory in terms of "hidden" variables*. Phys. Rev. **85** (1952) 166-179, *ibid* Phys. Rev. **85** (1952) 180-193
13. Valentini A., *Pilot-wave theory: an alternative approach to modern physics,* Cambridge University Press (2006).
14. Bricmont J., *Making sense of quantum mechanics*, Springer, Switzerland (2016).

15. Einstein A. Letter of 12 May 1952 to Max Born, in *The Born–Einstein Letters*, Macmillan, 1971, p. 192
16. Fuwa M., Takeda S., Zwierz M., Wiseman H. M., Furusawa A., *Experimental proof of nonlocal wavefunction collapse for a single particle using homodyne measurements* Nature Communication (2015) 7665
17. Bricmont J., *History of Quantum Mechanics or the Comedy of Errors*, arXiv:1703.00294v1, 1 Mar 2017.
18. MacDonald A. H., *Quantum Hall Effect: A Perspective*, Kluwer Academic Publishers (1989)
19. Prange R. E., Girvin S. M. (Eds.), *The Quantum Hall Effect*, Springer-Verlag, New York (1990).
20. Das Sarma S., Pinczuk A. (Eds.), *Perspectives in Quantum Hall Effects*, Wiley, New York (1997).
21. Weis J., von Klitzing K., *Metrology and microscopic picture of the integer quantum Hall effect*. Phil. Trans. R. Soc., 369 (2011), p. 3954
22. von Klitzing K., *The quantized Hall effect*, Rev. Mod. Phys., 58 (1986), p. 519-531
23. Haug R. J., *Edge-state transport and its experimental consequences in high magnetic fields*, Semicond. Sci. Technol., 8 (1993), p. 131-153
24. Halperin B. I., *Quantized Hall conductance, current-carrying edge states, and the existence of extended states in a two-dimensional disordered potential*, Phys. Rev. B, 25 (1982), p. 2185-2190
25. Emelyanov S. A. *From relativistic to quantum universe: Observation of a spatially-discontinuous particle dynamics beyond relativity*, Universe, **4** (2018) 75
26. Brosig S., Ensslin K., Warburton R. J., Nguyen C., Brar B., Thomas M., and Kroemer H., *Zero-field spin splitting in InAs-AlSb quantum wells revisited*, Phys. Rev. B 60 (1999), p. R13989-R13992
27. Bychkov Y. A., Rashba E. I., *Oscillatory effects and the magnetic susceptibility of carriers in inversion layers*, J. Phys. C, 17 (1984), p. 6039-6045
28. Bihlmayer G., Rader O., Winkler R., *Focus on the Rashba effect,* New J. Phys. 17 (2015), p. 050202
29. Gorbatsevich A. A., Kapaev V. V., Kopaev Yu. V., *Asymmetric nanostructures in a magnetic field*, JETP Lett., 57 (1993), p. 580-585
30. Gorbatsevich A. A., Kapaev V. V., Kopaev Yu. V., *Magnetoelectric phenomena in nanoelectronics*, Ferroelectrics, 161 (1994), p. 303-310
31. Bohm D., Hiley B. J., *The undivided universe: an ontological interpretation of quantum theory*, London Routledge (1993).
32. Mermin D., *Is the moon there when nobody looks? Reality and the quantum theory*, Physics Today, **38** April (1985) p. 38.
33. Landau L., Lifshitz E., *Quantum Mechanics, vol. 3, A Course of Theoretical Physics*, Pergamon Press, Oxford (1965) p. 2
34. Einstein A., Letter to Erwin Schrödinger (May 31, 1928), in K. Przibram (ed.), *Letters on Wave Mechanics, Schrödinger, Planck, Einstein, Lorentz*, Philosophical Library, NY (1967).
35. Feynman R. P., *Simulating physics with computers*, Int. J. Theor. Phys. **21** (1982) p. 471.
36. Ivchenko E. L., *Optical spectroscopy of semiconductor nanostructures*, Alpha Science Int., Harrow (2005).
37. Altarelli M., Maan J. S., Chang L. L., Esaki L., *Electronic states and quantum Hall effect in GaSb-InAs-GaSb quantum wells*, Phys. Rev. B, 35 (1987), p. 9867-9870
38. Ganichev S. D., Prettl W., *Intense terahertz excitation of semiconductors*, in Series on Semiconductor Science and Technology, vol. 14, Oxford University Press (2006).